\documentclass[aps,prb,showpacs,groupedaddress,amssymb,twocolumn]{revtex4} 
\usepackage{bm,amsmath} 
\usepackage{epsfig} 
\usepackage{txfonts} 
\begin{document} 
 
\title{Magnetic-moment oscillations in a quantum Hall ring} 
 
\author{Lachezar S. Georgiev} 
\affiliation{Institute for Nuclear Research and Nuclear Energy, 
72 Tsarigradsko Chaussee, 1784 Sofia, Bulgaria} 
 
\author{Michael R. Geller} 
 
\affiliation{Department of Physics and Astronomy, University of Georgia,  
Athens, Georgia 30602-2451} 
 
\date{April 28, 2004} 
 
\begin{abstract} 
We predict non-mesoscopic oscillations in the orbital magnetic moment of  
a thin semiconductor ring in the quantum Hall effect regime. These  
oscillations, which occur as a function of magnetic field because of a  
competition between paramagnetic and diamagnetic currents in the  
ring, are a direct probe of the equilibrium current distribution in the  
nonuniform quantum Hall fluid. The amplitude of the oscillating moment in a  
thin ring with major radius $R$ and minor radius (half width) $a$ is 
approximately $a R e  \omega_{\rm c} /c$, where $\omega_{\rm c}$ is the 
cyclotron frequency. 
\end{abstract} 
 
 
\maketitle 
\clearpage 
 
\section{introduction} 
 
There has been considerable interest in the orbital magnetic response of 
mesoscopic normal-metal rings to an applied magnetic field. The principal 
focus 
has been on the microscopic origin of the equilibrium persistent current $I$, 
which leads to a measurable magnetic moment $\pi R^2 I/c$ in a thin ring of 
radius $R$. The persistent current $I$ is mesoscopic in origin\cite{Imry} and  
vanishes in the macroscopic limit in accordance with Bloch's  
theorem.\cite{Vignale} However, because the current in a ring is actually  
{\it distributed}, the orbital magnetic moment 
\begin{equation} 
\mu \equiv {\pi \over c} \int\limits_0^\infty dr \ r^2 j(r), 
\label{magnetic moment definition} 
\end{equation} 
may be finite even if $I = \int_0^\infty dr \ j(r) $ 
vanishes. Here $j(r)$ is the azimuthal component of the equilibrium current 
density, and a two-dimensional electron system is assumed.  
 
This multipole-moment effect is negligible in metal rings 
at laboratory magnetic field strengths or in semiconductor rings in weak  
fields, because $j(r)$ is itself small under these conditions. However, the  
equilibrium current density in the quantum Hall regime generally consists of  
strips or channels of distributed current, which alternate in sign, and which  
have universal (confining potential independent) integrated strengths of the 
order of $e \omega_{\rm c}$  
\cite{Geller and Vignale,Geller and Vignale CDFT}. Therefore, although $I$ 
vanishes for a macroscopic  
ring in the quantum Hall regime, $\mu$ may be quite large. Furthermore, we  
shall show that because of a competition between paramagnetic and  
diamagnetic currents in the ring, the orbital magnetic moment oscillates with  
applied magnetic field. These oscillations, which are generally distinct from  
the de Haas--van Alphen or any mesoscopic oscillations, and which may be used  
to probe the equilibrium current distribution in the confined quantum Hall  
fluid, are the subject of this paper. 

\begin{figure}
\centering 
\caption{Two-dimensional electron gas in ring geometry.
\label{ring figure}} 
\epsfig{file=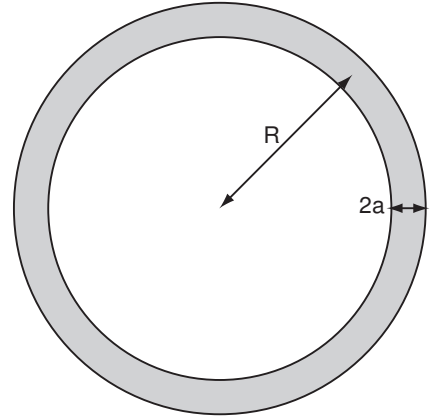,height=5.5cm} 
\end{figure} 

There is an extensive literature on finite-size orbital magnetic effects
in conductors,\cite{Denton, Robnik, Ruitenbeek, Zhu and Wang} 
and Sivan and Imry \cite{Sivan and Imry} have discussed the de 
Haas--van Alphen oscillations in small (simply connected) quantum 
dots. The limit of a two-dimensional electron gas in a strong perpendicular 
magnetic field and a parabolic confining potential has been solved exactly 
by Harrison.\cite{Harrison} However, we are not aware of any previous
work on the problem considered here.

In the next section, we discuss the equilibrium current distribution in a 
ring in the quantum Hall effect regime, and we present results of a  
self-consistent Hartree calculation of the oscillating magnetic moment. In  
Section III, we show that in a particular limit, the magnetic-moment  
oscillations become that of the de Haas--van Alphen effect for a  
two-dimensional electron gas in the area defined by the bulk of the ring.  
Section IV contains a discussion of our results. 
 
\section{hartree theory of the magnetic-moment oscillations} 
 
We shall assume that a two-dimensional electron gas is confined to a  
semiconductor ring of major radius $R$ and minor radius (half width) $a$ 
by a fixed positive 
background charge density of magnitude $n_0$. The inner radius of the  
two-dimensional ring is $R-a$ and the outer radius is $R+a$, as 
illustrated in Fig.~\ref{ring figure}. A uniform 
magnetic field of strength $B$ is oriented perpendicular to the 
two-dimensional electron gas in the $z$ direction.  
 
The origin of the magnetic-moment oscillations is best understood within a  
self-consistent Hartree approximation. The effects of exchange and correlation 
are discussed at the end of this paper.  
 
The theory of the equilibrium current distribution in the disorder-free, 
slowly confined quantum Hall fluid, has been developed in 
Refs.~\onlinecite{Geller and Vignale} and 
\onlinecite{Geller and Vignale CDFT}. In the ring geometry considered  
here, the current flows in the azimuthal direction. The azimuthal component 
of the current density, a function of the radial coordinate $r$ only, may 
be written (ignoring exchange and correlation effects) as 
\begin{equation} 
j(r) = j_{\rm edge}(r) + j_{\rm bulk}(r),
\label{current decomposition} 
\end{equation}  
where  
\begin{equation} 
j_{\rm edge} = {e \hbar \over m^*} \sum_{i=0}^\infty
 \left(i + {\textstyle{1 \over 2}} \right) \, 
{\partial n_i \over \partial r}, 
\label{edge} 
\end{equation} 
and 
\begin{equation} 
j_{\rm bulk} = {e n \over m^* \omega_{\rm c}}  \,  {\partial V \over  
\partial r}. 
\label{bulk} 
\end{equation} 
Here $m^*$ is the electron effective mass ($m^{*}/m\approx 0.067$ for GaAs), 
$e$ is the magnitude of the electron 
charge, and $\omega_{\rm c} \equiv e B / m^* c$ is the cyclotron frequency.  
In Eq.~(\ref{edge}) $n_i(r)$ denotes the contribution to the equilibrium 
number density 
\begin{equation} 
n(r) =  \sum_{i=0}^\infty n_i(r)
\label{density}
\end{equation}
from the $i$-th Landau level, namely
\begin{equation} 
n_i(r) \equiv  {1\over 2 \pi \ell^2} 
\sum_{\sigma} n_{\rm F}\left[\hbar \omega_{\rm c}\left( i + 
{\textstyle{1\over 2}} + {\textstyle{1\over 2}}\gamma \sigma\right) + 
V(r) \right] . 
\label{partial density} 
\end{equation} 
Here $n_{\rm F}(\epsilon)=[\exp(\epsilon -\mu)/k_{\rm B}T + 1]^{-1}$ 
is the Fermi distribution function, 
\begin{equation}
\gamma \equiv {g \mu_{\rm B} B \over \hbar \omega_{\rm c}} = 
{g m^* \over 2 m_{\rm e}} 
\label{spin splitting}
\end{equation}
is the dimensionless spin splitting (with $g$ the magnitude of the effective 
Land{\' e} $g$-factor of the host semiconductor and $\mu_{\rm B}$ 
the Bohr magneton),  $\ell \equiv \sqrt{\hbar c / e B}$ is the magnetic length,
and the summation in Eq.~(\ref{partial density}) is over spin components 
$\sigma = \pm 1$. 
Instead of using the bare $g$-factor, which is about 0.44 in GaAs,
we have used the experimentally measured exchange-enhanced value, 
which is about\cite{g7.3} $g\approx 7.3$ (see also Ref.~\onlinecite{g5.2} 
where a slightly smaller $g$-factor is observed). 
An important technical point in our computation is that the $g$-factor is
the same for all Landau levels. This has been confirmed in an 
experiment\cite{g5.2} showing that the exchange-enhanced $g$-factor 
is independent of the magnetic field despite the intuitive expectation that 
it should have a $B^{-1/2}$ dependence (see Fig.~4 in Ref.~\onlinecite{g5.2}).
The confining potential $V(r)$ in Eq.~(\ref{partial density}) consists of 
an external 
potential, produced by the fixed positive background  charge together with 
the self-consistent Hartree potential from the electron 
gas.\cite{confinement footnote} 
The expressions above are valid provided $\ell V' \ll \hbar \omega_{\rm c}$ 
and $k_{\rm B} T \ll \hbar \omega_{\rm c}$.
 
%%%%%%%%%%%%%%%%%%%%%%% 

Assuming cylindrical symmetry, and a positive background charge given by 
$n_{\rm b}(r),$
the Hartree potential energy in the ring is
\begin{equation}\label{pot}
V(r) = {4 e^2 \over \kappa} \int\limits_0^\infty \! dr' \, {r' \over r + r'} \, K\left( {4 r r' \over (r+r')^2} \right) 
\bigg[ n(r')-n_{\rm b}(r') \bigg],
\end{equation}
where
\begin{equation} \label{K}
K(m) \equiv \int\limits_0^{\pi/2} \frac{d\theta}{\sqrt{1-m \sin^2\theta}} 
\end{equation} 
is the complete elliptic integral of the first kind,\cite{abramowitz-stegun}
(see Appendix~\ref{app:elliptic} 
for more details) and $\kappa$ is the dielectric constant.
%%%%%%%%%%%%%%%%%%%%%%% 
\begin{figure}[htb] 
\centering 
\caption{The electron density for $\nu=2$ (solid line) plotted together with 
 the background charge density (dotted line) in units of  
$n_0=(2\pi r_0^2)^{-1}$ for a two-dimensional ring of major radius $R=40a$ at 
temperature $T=0.005 \ e^2/\kappa r_0 k_{\rm B}$. 
The inset shows in more detail the inner-edge region where the $\nu=1$ 
incompressible strip is formed.
\label{density figure}} 
\epsfig{file=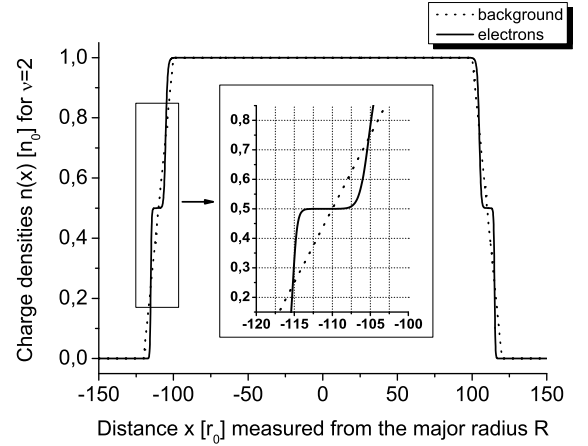,height=6.8cm} 
\end{figure} 
%%%%%%%%%%%%%%%%%%%%%% 

\begin{figure}[htb] 
\centering 
\caption{The current densities for $\nu=2$ (solid line)  and   
$\nu=5/2$ (dotted line) in units of $j_0=e\hbar/2\pi m^{*}r_0^3$ for a 
two-dimensional  ring of major radius $R=40a$ at temperature 
$T=0.005 \ e^2/\kappa r_0 k_{\rm B}$. 
\label{current figure}} 
\epsfig{file=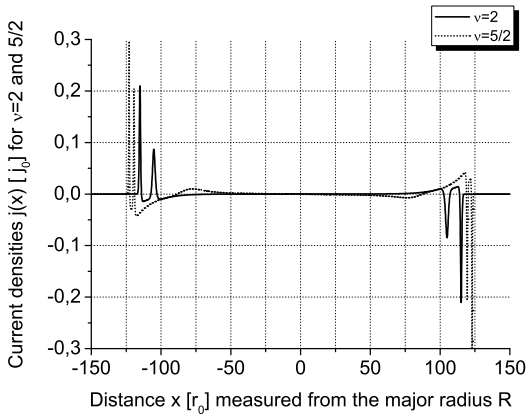,height=6.3cm} 
\end{figure}

In Figs.~\ref{density figure} and \ref{current figure} we present results of  
a self-consistent Hartree calculation of the low-temperature density and  
current 
distributions in the ring, as a function of $x \equiv r-R$ (the radial  
coordinate measured from the major radius of the ring). The dashed line in  
Fig.~\ref{density figure} shows the assumed positive background charge density 
of magnitude $n_0$ and width $2 a$, where $a = 100 r_0$. Here $r_0$ is the  
average interparticle spacing in the bulk of the ring, defined by $n_0 =  
(2 \pi r_0^2)^{-1}$. The background charge drops linearly to zero over a  
distance of $20 r_0$, leading to the trapezoid-shaped cross-sectional  
distribution shown. The corresponding low-temperature electron density 
for a two-dimensional ring of major radius $R=40 a$ is  
shown in the solid curve of Fig.~\ref{density figure}. The magnetic field  
strength has been chosen so that two Landau levels are filled ($\nu = 2$,  
where $\nu \equiv 2 \pi \ell^2 n_0 $) in the bulk of the ring. Wide  
incompressible strips are formed at filling factor $\nu = 1$ at the inner and  
outer edges of the ring. 
 
The azimuthal current density at this same magnetic field strength is shown in 
the solid curve of Fig.~\ref{current figure} in units of $j_0 \equiv e \hbar / 
2 \pi m^* r_0^3$.  
%%%%%%%%%%%%%%%%%%%%%% 
%%%%%%%%%%%%%%%%%%%%%% 
The qualitative features of the current distribution may be  
understood from Eqs.~(\ref{edge}) and (\ref{bulk}). At the inner edge of the 
ring, $x \approx -116 r_0$, the density increases from zero to $n_0/2$. 
Because the compressibility in this edge region is large, the screening is 
nearly perfect and the self-consistent confining potential is approximately 
uniform.  
Therefore, the current density is dominated by the edge contribution  
(\ref{edge}). The current is positive because $n(r)$ is increasing with  
radius. At the boundary of the innermost edge region, $x \approx -114 r_0$, 
the current has a node. In the $\nu =1$ incompressible region between 
$x \approx -114 r_0$ and $-107 r_0$, the density is uniform and the 
potential decreases by approximately $\hbar \omega_{\rm c}$. In this region 
the bulk contribution (\ref{bulk}) is dominant, and the current is negative 
because $\partial V / \partial r$ is negative. At $x \approx -107r_0$ the 
incompressible region ends and there is another node in the current. 
Another edge region is then encountered as the density approaches its bulk 
value of $n_0$. 
 
The current density at the outer edge of the ring follows similarly: There are 
two edge regions where the density decreases by $(2 \pi \ell^2)^{-1}$ and  
where the sign of the current is negative, separated  by a $\nu = 1$  
incompressible bulk region where the current is positive. We note 
that the integrated current in each channel is universal, independent of the  
shape of the channel and the details of the confining potential.\cite{Geller  
and Vignale}

It is tempting to regard the edge current (\ref{edge}) as being diamagnetic  
and the bulk current (\ref{bulk}) as paramagnetic, but this is not correct.  
Figure \ref{current figure} shows that the edge current---in the sense of our  
definition (\ref{edge})---at the outer edge of the ring is in fact  
diamagnetic, but that near the inner edge of the ring it is paramagnetic  
instead. 

At this point it is worth stressing the asymmetry between the odd-integer 
and even-integer filling factors $\nu$. 
It is well known that for the quantum Hall states with odd-integer $\nu$
the Fermi level resides in the Zeeman gap of size 
$\gamma \hbar\omega_{\rm c}$ within the uppermost Landau level, 
where $\gamma$ is defined in Eq.~(\ref{spin splitting}), while for 
the even-integer $\nu$ it lies in the cyclotron gap of size
$\hbar\omega_{\rm c}$. 
%%%%%%%%%%%%%%%%%%%%%%% 
\begin{figure}[htb] 
\centering 
\caption{The electron density for $\nu=3$ (solid line) plotted together with 
 the background charge density (dotted line) in units of  
$n_0=(2\pi r_0^2)^{-1}$ for a two-dimensional ring of major radius $R=40a$ at 
temperature $T=0.005 \ e^2/\kappa r_0 k_{\rm B}$. 
\label{density3}} 
\epsfig{file=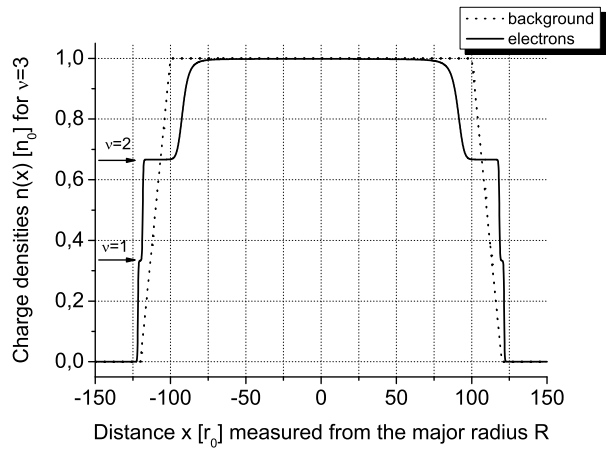,height=6.8cm} 
\end{figure} 
%%%%%%%%%%%%%%%%%%%%%% 
In our calculations, the spin-splitting is about 25\% of the cyclotron 
energy.
Because the width of the incompressible channel  is proportional  to the 
corresponding energy gap, the incompressible strips 
formed at odd-integer $\nu$ are notably narrower than the even-integer ones,
as evident in Fig.~\ref{density3}. 
The horizontal arrows in Fig.~\ref{density3} point to the positions 
of the incompressible strips corresponding to filling factors 
$\nu=1$ (at $n/n_0=1/3$) and $\nu=2$ (at $n/n_0=2/3$).
 
As the magnetic field is changed, the filling factor in the bulk of the ring  
changes. It is useful to define a dimensionless magnetic field strength 
\begin{equation} 
\beta \equiv {\hbar \omega_{\rm c} \over e^2/\kappa r_0}, 
\label{dimensionless field strength} 
\end{equation} 
where, as before, $\kappa$ is the dielectric constant of the semiconductor. The density of 
one filled Landau level is given by $(2 \pi \ell^2)^{-1} = r_{\rm s} \beta  
n_0$, where $r_{\rm s} \equiv r_0/ a_{\rm B}$ is the dimensionless  
interparticle spacing and $a_{\rm B} \equiv \hbar^2 \kappa / m^* e^2$ is the  
effective Bohr radius in the semiconductor. The density in Fig.~\ref{density  
figure} and current in the solid curve of Fig.~\ref{current figure} were  
calculated with the parameter values $r_{\rm s} = 1$ and $\beta = 1/2$. 
 
The dotted curve of Fig.~\ref{current figure} shows the current distribution 
after the magnetic field has been reduced to a strength corresponding to 
$\beta = 2/5$, so that $\nu = 5/2$ in the bulk of the ring. In this case, 
there 
are three edge channels and two incompressible strips at the inner and outer  
edges of the ring. Note that the current distribution for $\beta = 1/2$ and  
$\beta = 2/5$ are almost everywhere opposite in sign (after an unimportant
shift in $x$). This sign reversal,  
which clearly leads to oscillations in the orbital magnetic moment of the  
ring, occurs because of a combination of two factors: The competition between  
the edge current (\ref{edge}) and bulk current (\ref{bulk}) leads to a pattern 
of strips or channels of current with alternating sign, and the  
electron-electron interaction expels most of the current from the bulk to the 
inner and outer edges of the ring.
 
\begin{figure}[htb] 
\centering 
\caption{The magnetic moment as a function of the filling  
factor in units of $\mu_0$ for a two-dimensional ring of major radius 
$R=40a$ for temperature $T=0.005 \ e^2/\kappa r_0 k_{\rm B}$. 
The dashed line shows the oscillating magnetic moment of the noninteracting 
electron gas at $T=0$. 
\label{magnetic moment figure}} 
\epsfig{file=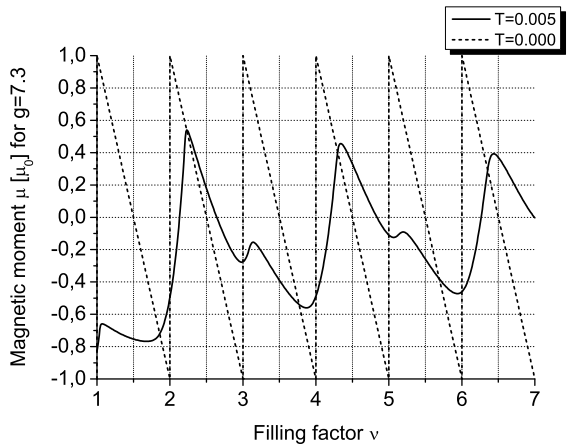,height=6.7cm} 
\end{figure}

The orbital magnetic moment of the ring is straightforward to compute from  
(\ref{magnetic moment definition}), and is plotted in the solid curve of  
Fig.~\ref{magnetic moment figure} in units of  
\begin{equation} 
\mu_0 \equiv 4 \pi R a n_0 \mu_{\rm B}^* ,  
\label{moment unit} 
\end{equation} 
where $\mu_{\rm B}^* \equiv e \hbar / 2 m^* c$ is the effective Bohr magneton. 
The magnetic moment is plotted as a function of $\nu$, the filling factor in  
the bulk of the ring, and $r_{\rm s} = 1$ as before. 
Note that the asymmetry between the odd-integer and even-integer $\nu$ 
discussed above leads to a partial suppression of the magnetization 
oscillation at odd-integer $\nu$.
The dashed line of  
Fig.~\ref{magnetic moment figure} shows the magnetic moment in the limit $R  
\gg a \gg \ell$ and $T = 0$, which is calculated analytically in the next  
section.  The deviation of the solid curve from the dashed one is a consequence
of both finite-size effects and finite temperature.
The solid curve on Fig.~\ref{magnetic moment figure} is essentially the same 
as that in Fig.~2 of Ref.~\onlinecite{magnetization-exp}. 
The magnetic field distance 
between two adjacent minima of the magnetization in Fig.~2 in 
Ref.~\onlinecite{magnetization-exp} 
corresponds to a distance in $\nu$ approximately equal to 2 which is 
in very good agreement with Fig.~\ref{magnetic moment figure} bellow.
%%%%%%%%%%%%%%%%%%%%%% 
%%%%%%%%%%%%%%%%%%%%%% 

\begin{figure}[htb] 
\centering 
\caption{The magnetic moment (at temperature 
$T=0.005 \ e^2/\kappa r_0 k_{\rm B}$) as a function of the total magnetic field
$B_{\rm tot}$ computed for the sample T412 of 
Ref.~\onlinecite{magnetization-exp} (after illumination) in 
units of $\mu_0$ for a two-dimensional ring of major radius 
$R=40a$. The arrows show the magnetic fields at which the corresponding 
filling factors appear. \label{mu-B} } 
\epsfig{file=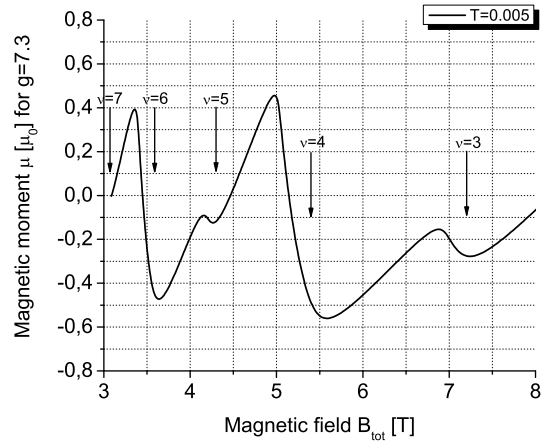,height=6.7cm} 
\end{figure}

In addition to the plot of the magnetic-moment oscillations as a 
function of the filling factor $\nu$ given in 
Fig.~\ref{magnetic moment figure}, 
it is interesting to plot the magnetic moment as a function of the magnetic 
field. This is given in Fig.~\ref{mu-B} for a system 
with $n_0=4.9\times 10^{15} m^{-2} \! $, 
corresponding to sample T412 of Ref.~\onlinecite{magnetization-exp} 
(after illumination),
where we find excellent agreement with that experiment (see Fig.~3 of 
Ref.~\onlinecite{magnetization-exp}). 

%%%%%%%%%%%%%%%%%%%%%% 
%%%%%%%%%%%%%%%%%%%%%%
 
\section{moment oscillations in the de haas--van alphen limit}
\label{sect:dHvA}  
 
In this section we calculate the orbital magnetic moment for the  
special case of $R \gg a \gg \ell$, a limit where an exact analytic solution  
is possible. Recall that $2a$ is the width of the electron gas in a ring of  
radius $R$, and $\ell$ is the magnetic length. We shall show that in this  
limit the magnetic-moment oscillations become that of the de Haas--van Alphen  
effect for a two-dimensional electron gas in the area defined by the bulk of  
the ring. 
 
In terms of $x \equiv r - R$, the magnetic moment (\ref{magnetic moment  
definition}) may be written as 
\begin{equation} 
\mu = {\pi \over c} \int\limits_{-\infty}^\infty dx \ (R+x)^2 \ j(x) , 
\end{equation} 
where the fact that the current density vanishes outside the ring has been  
used to extend the limits of integration. Because $j(x)$ is an odd function of 
$x$ in a macroscopic ring  in the $R \gg a$ limit [see Eq.~(\ref{symmetry}) in 
Appendix~\ref{app:elliptic}], we have 
\begin{equation}\label{j_odd} 
\mu = {4 \pi R \over c} \int\limits_{0}^\infty dx \ x \ j(x) . 
\end{equation} 
This expression shows that $\mu$ is proportional to the first moment of the  
current distribution shown in Fig.~\ref{current figure}, which itself  
oscillates as a function of magnetic field. 
 
Because the current density is concentrated at the inner and outer edges of 
the 
ring on a length scale much smaller than $a$, 
\begin{equation} 
\int\limits_{0}^\infty dx \ x \ j(x)  
\approx a \! \int\limits_{0}^\infty dx \ j(x).  
\end{equation} 
Finally, the relation ${\bf j}({\bf r}) = c \nabla \times {\bf M}({\bf r})$ 
between current and local magnetization shows that 
\begin{equation} 
\int\limits_{0}^\infty dx \ j(x) = c \ \! M(\nu), 
\end{equation} 
where $M(\nu)$ is the $z$ component of the equilibrium orbital magnetization 
of a uniform two-dimensional electron gas with filling factor $\nu$. 

Therefore, in the de  
Haas--van Alphen limit, 
\begin{equation} 
\mu = 4 \pi R a M(\nu).  
\label{dH-vA limit} 
\end{equation}
The expression (\ref{dH-vA limit}) shows that the magnetic moment in the $R  
\gg a \gg \ell$ limit is equal to that of a uniform electron gas of filling  
factor $\nu$ and area $4 \pi R a$, the area of the ring of width $2a$ and  
radius $R$. The magnetic moment in this limit, which may also be written as 
\begin{equation} 
\mu = \mu_0 \  {M(\nu) \over \mu_{\rm B}^* n_0} , 
\end{equation} 
is shown at zero temperature in the dashed curve of Fig.~\ref{magnetic moment  
figure}.   
 
\section{discussion} 
 
In this paper, we have shown that the equilibrium current density in a  
confined quantum Hall fluid, which generally consists of a series of strips or 
channels of distributed current with alternating signs, leads to a magnetic  
moment in a non-mesoscopic ring that oscillates as a function of magnetic  
field. In a certain limit, namely $R \gg a \gg \ell$, the oscillations may be  
regarded as de Haas--van Alphen oscillations of the Hall fluid in the bulk of  
the ring, even though the current density in this limit is actually  
concentrated at the inner and outer edges.  
 
The most natural interpretation of the magnetic-moment oscillations reported  
here is one in terms of the multipole moments of the equilibrium current  
distribution. The magnitude of the net current flowing at the edge of $\nu$  
filled Landau levels is of the order of $\nu e \omega_{\rm c}$. This current  
contributes approximately $R^2 \nu e \omega_{\rm c}/ c$ to the orbital  
magnetic moment. However, the equilibrium currents flowing at the inner and  
outer edges of a macroscopic ring cancel, so the zeroth-order radial moment of 
the azimuthal current distribution, $\int_{-\infty}^\infty dx \ j(x)$,  
vanishes. However, the first radial moment, $\int_{-\infty}^\infty dx \ x \,  
j(x)$, does not vanish. The resulting magnetic moment is of the order of  
$\mu_0$, defined in (\ref{moment unit}), which may be equivalently 
rewritten as 
\begin{equation} 
\mu_0 = \bigg({a \over R} \bigg) \, R^2 \nu e \omega_{\rm c}/c, 
\end{equation} 
making evident its multipole-moment origin. 
 
Finally, we note that the spin polarization contributes to
the total the magnetic moment of the ring an amount  
\begin{equation} 
\mu_{\rm spin} = - {\textstyle{1 \over 2}} g \mu_{\rm B} (N_\uparrow -  
N_\downarrow), 
\end{equation} 
where $N_\sigma$ is the number of electrons with spin $\sigma$.
The principal effect of exchange is to enhance the bare Zeeman spin 
splitting, which we have already accounted for phenomenologically by using
$g$-factors deduced by experiments. The most important  
effect of correlation on the zero-temperature chemical potential  
and orbital magnetization of the uniform Hall fluid is to introduce additional 
discontinuities at certain fractional filling factors. This additional  
structure leads to incompressible strips and associated bulk currents at  
fractional filling factors in the nonuniform Hall fluid. It is clear from the  
analysis of Section III that in the $R \gg a \gg \ell$ limit, the orbital  
magnetic moment will reflect the actual orbital magnetization of the  
interacting electron gas. For other values of $R$ and $a$, we expect a more  
complex oscillatory response. Measurements of the magnetic moment of a  
semiconductor ring in the quantum Hall regime would provide useful information 
about the equilibrium current distribution in the nonuniform Hall fluid.

\begin{acknowledgments} 
 
L.S.G. thanks  Boyan Obreshkov for useful discussions and Dimitar Bakalov 
for access to a powerful PC. L.S.G. has been partially supported by the  
FP5-EUCLID Network Program of the European Commission under Contract No. 
HPRN-CT-2002-00325. M.R.G. thanks Michael Harrison 
for useful discussions, and the National Science Foundation and Research 
Corporation for support under CAREER Grant No.~DMR-0093217 and for
a Cottrell Scholars Award, respectively.
 
\end{acknowledgments} 
 
\appendix 
%%%%%%%%%%%%%%%%%%%%%%%%%%%%%%%%%%%%%%%%%%%%% 
\section{The elliptic integral} 
\label{app:elliptic} 
%%%%%%%%%%%%%%%%%%%%%%%%%%%%%%%%%%%%%%%%%%%%% 
The electrostatic potential produced by the electric charge distribution 
$\rho({\bf r'})=e \, n(r')\delta(z')$ is explicitly independent of the  
polar angle $\phi$ and takes the form  
\[ 
V(r,z)=\frac{e^2}{\kappa}\int d^3 r' \frac{n(r')\delta(z')} 
{\sqrt{(\vec{r}-\vec{r}')^2+(z-z')^2}} 
\] 
where $\vec{r}$ and  $\vec{r}'$ are the projections of the 3D vectors  
${\bf r}$ and  ${\bf r}'$, respectively in the plane $z=0$. Let us choose the 
vector $\vec{r}$ to be in the direction of the $x$ axis so that 
$(\vec{r}-\vec{r}')^2=r^2+{r'}^2-2rr'\cos\phi'$ where $\phi'$ is the  
polar angle of  $\vec{r}'$. After integrating over $z'$   
(and setting $z=0$) we get 
\begin{equation}\label{V} 
V(r)=\frac{e^2}{\kappa}
\int\limits_0^\infty r' dr' n(r') \int\limits_{-\pi}^\pi   
\frac{d \phi'}{\sqrt{r^2+{r'}^2 -2rr' \cos \phi' } }. 
\end{equation} 
Noting that $\cos(-\phi')=\cos(\phi')$ and changing the integration variable 
$\phi'=\pi-2\theta$ we can express the integral over $\phi'$ in the above  
equation in terms of the complete elliptic integral (\ref{K}), namely,  
\begin{equation}\label{elliptic} 
\frac{4}{r+r'} K(m) \quad 
\mathrm{with}\quad m=\frac{4rr'}{(r+r')^2}. 
\end{equation} 
Substituting Eq.~(\ref{elliptic}) back into  
Eq.~(\ref{V}) one recovers the expression (\ref{pot}) that have been used 
in the self-consistent computation of the potential. 
 
It is worth noting that in the de Haas-van Alphen limit in 
Sect.~\ref{sect:dHvA}, in which $R\gg a$, the electrostatic potential 
reproduces that of a thin infinite strip. The point is that 
$R-a\leq r, \, r' \leq R+a$ and  in this case $r\approx r'\approx R$
so that $m\approx 1$ as seen from Eq.~(\ref{elliptic}). However the  
elliptic integral (\ref{K}) has a logarithmic asymptotics 
\cite{abramowitz-stegun} for $m \to 1$
\[
\lim_{m\to 1} 
\left[ K(m) - \ln\left(\frac{4}{\sqrt{1-m}}\right) \right]=0,
\]
where in our case $1-m=(r-r')^2/4R^2$. Setting $r= R+x$ and $r'=R+x'$  
we obtain
\begin{equation}\label{log}
V_{\rm dHvA}(x) = - {2 e^2 \over \kappa} \int\limits_{-\infty}^\infty \! dx' 
\, \ln\left|x-x' \right| 
\bigg[ n(x')-n_{\rm b}(x') \bigg],
\end{equation}
where we have used $\int \! dx \ n(x)=\int \! dx \ n_{\rm b} (x)$.
Equation~(\ref{log}) is exactly the potential of electrons in a 2D 
infinite strip with number density $n(x)$ and background charge 
density $-n_{\rm b}(x)$.
It follows from Eqs.~(\ref{edge}),  (\ref{bulk}), 
(\ref{partial density}) and  (\ref{log}) that in the de Haas--van Alphen 
limit the potential and charge densities are even functions of $x$ 
while the current densities are odd 
\begin{equation}\label{symmetry}
V(-x)=V(x), \quad n(-x)=n(x), \quad {\rm and} \quad j(-x)=-j(x)
\end{equation}
which has been used in the derivation of Eq.~(\ref{j_odd}).
This symmetry is not present in a general two-dimensional electron ring.
 
%%%%%%%%%%%%%%%%%%%%%%%%%%


\begin{thebibliography}{99} 
 
\bibitem{Imry} Y. Imry, {\it Introduction to Mesoscopic Physics} 
(Oxford University Press, Oxford, 1997). 
 
\bibitem{Vignale} G. Vignale, Phys. Rev. B {\bf 51}, 2612 (1995). 
 
\bibitem{Geller and Vignale} M. R. Geller and G. Vignale, Phys. Rev. B  
{\bf 50}, 11714 (1994).

\bibitem{Geller and Vignale CDFT} M. R. Geller and G. Vignale, Phys. Rev. B  
{\bf 52}, 14137 (1995). 

\bibitem{Denton} R. V. Denton, Z. Phys. {\bf 265}, 119 (1973).

\bibitem{Robnik} M. Robnik, J. Phys. A {\bf 19}, 3619 (1986).

\bibitem{Ruitenbeek} J. M. van Ruitenbeek, Z. Phys. D {\bf 19}, 247 (1991).

\bibitem{Zhu and Wang} J.-X. Zhu and Z. D. Wang, Phys. Lett. A {\bf 203}, 
144 (1995).

\bibitem{Sivan and Imry} U. Sivan and Y. Imry, Phys. Rev. Lett. {\bf 61},  
1001 (1988). 
 
\bibitem{Harrison} M. J. Harrison, Phys. Rev. B {\bf 45}, 3815 (1992). 

\bibitem{g7.3} A. Usher, R. J. Nicholas, J. J. Harris and C. T. Foxon,
Phys. Rev. {\bf B 41}, 1129 (1990).

\bibitem{g5.2} V. T. Dolgopolov, A. A. Shashkin, A. V. Aristov, D. Schmerek, 
W. Hansen, J. P. Kotthaus and M. Holland, Phys. Rev. Lett. {\bf 79}, 
729 (1997).

\bibitem{confinement footnote} As is often the case with self-consistent 
Hartree
and Hartree-Fock calculations in the quantum Hall effect regime, we find
it necessary to include an additional hard-wall confining potential to V.
Its precise form does not appreciably affect our results.
  
\bibitem{abramowitz-stegun} {\it Handbook of Mathematical Functions}, 
edited by M. Abramowitz and I. Stegun (Dover, New York, 1972). 
 
\bibitem{magnetization-exp} A. Usher, M. Zhu, A. J. Matthews, A. Potts, 
M. Elliott, W. G. Herrenden-Harker, D. A. Ritchie and M. Y. Simmons,
Physica {\bf E 22}, 741 (2004).


\end{thebibliography}
\end{document}